# Moiré Topological Magnetism Twist-Engineered from 2D Spin Spirals

Zhonglin He,[1,2] Kaiying Dou,[1] Wenhui Du,[1] Ying Dai,[1] Evgeny Y. Tsymbal,[2,*] and Yandong Ma[1,*]

[1]*School of Physics, State Key Laboratory of Crystal Materials, Shandong University, Jinan 250100, China*

[2]*Department of Physics and Astronomy & Nebraska Center for Materials and Nanoscience, University of Nebraska, Lincoln, Nebraska 68588-0299, USA*

[*]Corresponding authors: tsymbal@unl.edu; yandong.ma@sdu.edu.cn

## Abstract

Topological magnetism, characterized by topologically protected spin textures, offers rich physics and transformative prospects for spintronics. However, its stabilization typically demands external magnetic fields, preventing straightforward implementation. Here, we report a universal field-free approach for engineering 2D topologically-trivial spin spirals into topological magnetisms. This approach leverages twisted antiferromagnetic bilayers, where locked spin spirals in the two sublayers form spatially alternating ferromagnetic and antiferromagnetic domains upon twisting. These domains frustrate the uniform antiferromagnetic interlayer exchange, spontaneously stabilizing moiré topological magnetisms without external fields. Using first-principles and atomistic spin-model simulations, we validate this approach using bilayers $NiCl_2$ and $NiBr_2$, as representative examples. For twisted $NiCl_2$, we predict topological spin states tunable by the twist angle, including isolated and high-order antiferromagnetic bimerons. For twisted $NiBr_2$, strong frustration yields trivial triple-**q** spin spirals, which transform into moiré topological magnetism with the application of vertical compressive strain. Our findings demonstrate that topologically non-trivial spin textures can be engineered from their trivial counterparts, thus providing a new paradigm for topological spintronics.

**Keywords**: Moiré topological magnetism, first-principles calculations, frustrated magnetism, two-dimensional lattice, twist engineering

# Introduction

Topological magnetism, characterizing topologically protected magnetic quasiparticles in real space, is one of the central topics in condensed matter physics and materials science [1-3]. It offers promising perspectives for information technology [4-6] and serves as a fertile platform for investigating fundamental phenomena such as the topological Hall effect [7], topological spin Hall effect [8], and skyrmion Hall effect [9]. The advent of two-dimensional (2D) magnetic materials [10-12] has further accelerated research in this field, attracting broad interest [13-16]. The emergence of topological magnetism in 2D systems stems from a delicate balance of competing magnetic interactions [17]. In this context, most 2D materials capable of hosting non-coplanar spin structures intrinsically stabilize trivial spin spirals rather than topological magnetism [18-20]. Current strategies to stabilize topological magnetism in such systems typically require additional external magnetic fields [21-23] posing significant constraints on practical device integration. There is thus an urgent need to explore new approaches for generating topological magnetism from 2D trivial spin spirals without external fields.

Moiré superlattices in 2D materials have served as a versatile platform for realizing new electronic phases (dubbed moiré electronics), as exemplified by the recent discoveries of topological insulators in twisted $MoTe_2$ [24,25] as well as unconventional superconductivity in twisted $WSe_2$ [26,27]. Inspired by moiré electronics, the magnetic counterpart, moiré magnets created by twisting magnetic atomic layers, is anticipated to host emergent magnetic phases distinct from those in natural 2D magnets [28,29]. Recent studies show that interlayer twist can induce non-coplanar spins in intrinsically collinear 2D magnetic systems [30,31] or modulate the physical properties of inherently 2D topological magnetic materials [32,33]. For instance, Antão et al. and Zhu et al. [34, 35] reported that in twisted bilayer, spatial variations in local point group symmetry of the moiré lattice can modulate the pre-existing topological properties [23, 36]. This raises the pivotal question of whether the twist can realize isolated moiré topological magnetisms in 2D systems originally hosting trivial spin spirals. Critically, no general strategy has yet been established to transform such trivial spirals into topologically protected moiré magnetism— particularly without external fields.

In this letter, we introduce a universal approach for generating and controlling moiré topological magnetism in 2D systems originally hosting trivial spin spirals. The idea is to stack such systems into antiferromagnetic (AFM) bilayers. Upon twisting, local overlap of the locked spin spirals in the two sublayers forms spatially alternating ferromagnetic (FM) and AFM domains. These domains frustrate with the uniform AFM interlayer exchange interaction, spontaneously stabilizing moiré topological magnetism without external magnetic field. By performing first-principles calculations and atomistic spin model simulations, we demonstrate this mechanism for bilayer Ni$X_2$, where $X$ = Cl or Br. For NiCl$_2$, we predict the emergence of topological spin states tunable by the twist angle, including isolated AFM bimerons, AFM merons, AFM antimerons, and high-order AFM bimerons. In NiBr$_2$, strong frustration stabilizes trivial AFM triple-**q** spirals, which evolves into topological magnetic textures under vertical compressive strain. These results establish a new paradigm for topological spintronics, where the topologically non-trivial spin textures can be twist-engineered from their trivial counterparts.

## Results and Discussion

To engineer spin spirals in 2D materials, we first consider a single-layer (SL) magnetic lattice to set the stage. Non-coplanar spin structures arise primarily from two mechanisms: the Dzyaloshinskii-Moriya interaction (DMI) induced by broken inversion symmetry, and magnetic frustration, which may be intrinsic to the material. The spin structures of a SL lattice can be described by the model Hamiltonian

$$H_{\text{intra}} = -\sum_{\langle i,j \rangle} \mathbf{S}_i \mathbf{J}_{1,ij} \mathbf{S}_j - \sum_{\langle\langle i,j \rangle\rangle} J_2 \mathbf{S}_i \cdot \mathbf{S}_j - \sum_{\langle\langle\langle i,j \rangle\rangle\rangle} J_3 \mathbf{S}_i \cdot \mathbf{S}_j - \sum_{\langle i \rangle} K(\mathbf{S}_i \cdot \hat{\mathbf{z}})^2. \quad (1)$$

Here, $\mathbf{S}_i$ denotes the spin vector at site $i$ (position $\mathbf{r}_i$). $\hat{\mathbf{z}}$ is the unit vector perpendicular to the plane of 2D magnetic lattices. The nearest-neighbor (NN) exchange interaction tensor $\mathbf{J}_{1,ij}$ decomposes into

$$\mathbf{J}_{1,ij} = \mathbf{J}^I_{1,ij} + \mathbf{J}^A_{1,ij} + \mathbf{J}^S_{1,ij}, \tag{2}$$

comprising isotropic $\mathbf{J}^I_{1,ij} = \frac{1}{3}\mathrm{Tr}(\mathbf{J}_{1,ij})\mathbf{I}$, antisymmetric DMI $\mathbf{J}^A_{1,ij} = \frac{1}{2}(\mathbf{J}_{1,ij} - \mathbf{J}^T_{1,ij})$, and anisotropic symmetric $\mathbf{J}^S_{1,ij} = \frac{1}{2}(\mathbf{J}_{1,ij} + \mathbf{J}^T_{1,ij}) - \mathbf{J}^I_{1,ij}$ terms. We include exchange interactions up to the third nearest neighbor within the $J_1$-$J_2$-$J_3$ model and magnetic anisotropy $K$. With tuning these parameters, spin spirals with different wavevector $\mathbf{q} = (q_x, q_y)$ can be obtained. For simplicity, we consider single-$\mathbf{q}$ spin spirals where we assume that $\mathbf{q} = (q_x, 0)$ is favored, so that

$$\mathbf{S}_i(\mathbf{r}_i) = \cos(\mathbf{q}\cdot\mathbf{r}_i + \phi)\hat{\mathbf{x}} + \sin(\mathbf{q}\cdot\mathbf{r}_i + \phi)\hat{\mathbf{z}}, \tag{3}$$

where $\hat{\mathbf{x}}$ is the unit vector along $x$ and phase $\phi$ describes translation of spin spirals (**Fig. S1** [45]).

We then stack two such SL lattices together to construct a bilayer lattice. Initially neglecting interlayer interactions, the bilayer spin Hamiltonian is

$$H = \sum_{l=t,b} H^l_{\mathrm{intra}}, \tag{4}$$

where $l = t, b$ labels the top and bottom layers, respectively. It is known that in a bilayer lattice, the interlayer exchange interaction is usually dominated with a uniform A-type AFM coupling regardless of the stacking pattern [37-39]. Following this principle, despite the absence of explicit interlayer coupling, we impose an inherent A-type AFM alignment. This results in AFM single-$\mathbf{q}$ spin spirals with $\mathbf{q} = (q_x, 0)$ in the AA-stacked bilayer lattice, as illustrated in **Fig. S1** [45]. The spin vector can be expressed by $\mathbf{S}^l_i(\mathbf{r}^l_i) = \cos(\mathbf{q}\cdot\mathbf{r}^l_i + \phi^l)\hat{\mathbf{x}} + \sin(\mathbf{q}\cdot\mathbf{r}^l_i + \phi^l)\hat{\mathbf{z}}$, and $\phi^l$ is set to be 0 and 180° for the top and bottom layers, respectively.

Upon rotating the top layer with a twist angle $\theta$, as illustrated in **Fig. 1(a)**, the local overlap between the locked spin spirals of the sublayers is cased, forming spatially alternating FM (red and blue areas) and AFM (purple areas) domains. The twisted top-layer spins now follow:

$$\mathbf{S}^t_i(\mathbf{r}^t_i) = \cos(\mathbf{q}\cdot\mathcal{R}(\theta)\mathbf{r}^t_i)\hat{\mathbf{x}} + \sin(\mathbf{q}\cdot\mathcal{R}(\theta)\mathbf{r}^t_i)\hat{\mathbf{z}}, \tag{5}$$

where $\mathcal{R}(\theta) = \begin{pmatrix} \cos\theta & -\sin\theta \\ \sin\theta & \cos\theta \end{pmatrix}$ is the in-plane rotation matrix. The bottom-layer spins remain

$$\mathbf{S}_i^b(\mathbf{r}_i^b) = -\cos(\mathbf{q}\cdot\mathbf{r}_i^b)\hat{\mathbf{x}} - \sin(\mathbf{q}\cdot\mathbf{r}_i^b)\hat{\mathbf{z}}. \tag{6}$$

Including interlayer interactions introduces an additional spin Hamiltonian term:

$$H_{\text{inter}} = -\sum_{\langle i,j \rangle} J_\perp(\mathbf{r}_i^t - \mathbf{r}_j^b)\mathbf{S}_i^t \cdot \mathbf{S}_j^b = -\sum_{\langle i,j \rangle} J_\perp(\mathbf{r}_{ij})\mathbf{S}_i^t \cdot \mathbf{S}_j^b, \tag{7}$$

where the distance-dependent interlayer coupling $J_\perp(\mathbf{r}_{ij}) = J_\perp(\mathbf{r}_i^t - \mathbf{r}_j^b) < 0$ enforces uniform AFM interlayer coupling. Based on the interlayer spin Hamiltonian and spin spirals in the two sublayers, we can obtain the interlayer exchange energy (see Sec. II of Supplemental Material [45]). **Fig. 1(b)** displays the interlayer exchange energy versus position along the interdomain midline A-B [see **Fig. 1(a)**], revealing characteristic oscillations that energetically suppress FM domains. Crucially, this spatial alternation between FM and AFM domains frustrates the uniform AFM interlayer coupling. Since topological magnetic quasiparticles typically originate from competing magnetic interactions, such frustration can enable the spontaneous stabilization of moiré topological magnetism (i.e., emergent topological magnetic quasiparticles within moiré lattices), offering a novel pathway to engineer topological magnetism from trivial spin spirals in 2D systems. Additionally, this scenario also applies to 2D materials with multi-**q** spin spirals.

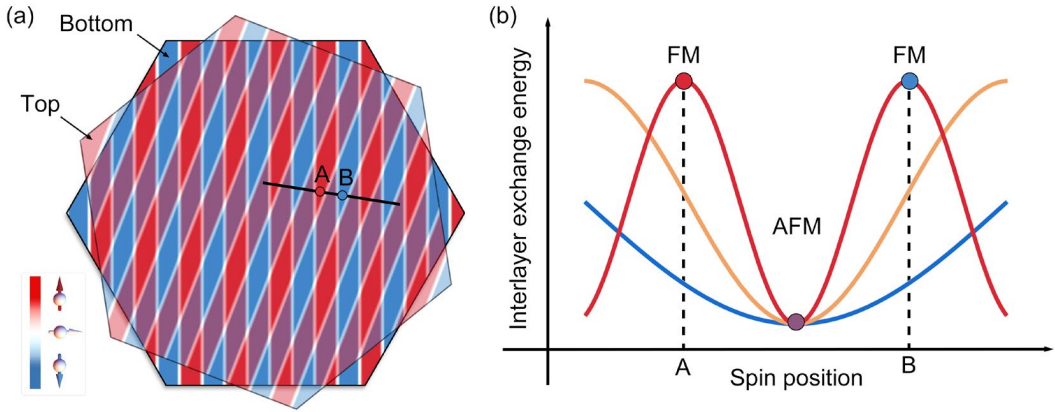

**Fig. 1**. (a) Schematic illustration of the spin spirals in twisted AFM bilayer in the absence of interlayer coupling. Out-of-plane spin components are denoted by colors. (b) Interlayer exchange

energy versus spin position along the interdomain midline A-B, as indicated in (a). Blue, orange and red lines in (b) represent interlayer exchange energies under twist angles of $\theta = 1.960°$, $4.040°$ and $9.430°$, respectively.

Two candidate materials to validate our proposed approach are Ni$X_2$ ($X$ = Cl, Br). SL Ni$X_2$ crystallizes in a triangular lattice (space group $P\bar{3}m1$; **Fig. S2** [45]), with lattice parameters $a_0$ listed in **Table S1**, agreeing well with the previous work [23]. The electronic configuration of isolated Ni atom is $3d^8 4s^2$. By donating two electrons to $X$ ligands, Ni adopts a +2 oxidation state in Ni$X_2$. Under an octahedral crystal field, the $d$ orbitals of Ni split into $t_{2g}$ and $e_g$ manifolds. The Ni atom exhibits an electric configuration of $t_{2g}^6 e_g^2$, which is expected to produce the magnetic moment of 2 $\mu_B$. As anticipated, our calculations reveal that monolayer Ni$X_2$ favors a spin-polarized state with a magnetic moment of 2 $\mu_B$ per unit cell, primarily localized at Ni atoms.

Next, we perform Density Functional Theory (DFT) calculations to investigate the magnetic interactions in SL Ni$X_2$ according to **Eq. (1)**. The four-state method is introduced to obtain exchange interactions $J$ and magnetic anisotropy $K$ (see Sec. III of Supplemental Material [45]). As listed in **Table. S1**, SL Ni$X_2$ exhibits FM NN exchange interaction ($J_1 > 0$) and AFM third NN isotropic exchange interaction ($J_3 < 0$), while the FM next NN isotropic exchange interaction is negligible. This hierarchy stems from the super-exchange mediated by delocalized $p$ states of $X$ atoms [23,40]. We find that the ratio of $J_3/J_1$ is –0.33 and –0.49 for NiCl$_2$ and NiBr$_2$, respectively, revealing the increasing exchange frustration across the ligand series. Due to inversion symmetry, DMI is forbidden in SL Ni$X_2$. The obtained negative values of $K$ suggest that they exhibit in-plane magnetization. To explore the spin textures in SL Ni$X_2$, we perform atomistic spin model simulations based on the spin Hamiltonian of **Eq. (1, 2)** and the Landau-Lifshitz-Gilbert (LLG)-Heun method implemented in the VAMPIRE package (see Sec. I of Supplemental Material [45]). As shown in **Fig. S3** [45], both systems exhibit trivial spin-spiral textures. Such trivial spin spirals are further confirmed by the spin structure factor [41]

$$S(\mathbf{q}) = (1/N^2) \sum_{\mu=x,y,z} \left\langle \left| \sum_i \mathbf{S}_i \cdot \hat{\mathbf{r}}_\mu \exp(-i\mathbf{q} \cdot \mathbf{r}_i) \right|^2 \right\rangle, \tag{8}$$

where $N$ and i represent the number of spins and imaginary unit, respectively. Whereas SL $NiCl_2$ exhibits multi-**q** spin spirals, SL $NiBr_2$ favors single-**q** spin spirals (**Fig. S3** [45]). This distinction is due to their different exchange frustration strengths, namely $J_3/J_1$ is larger for $NiBr_2$ than for $NiCl_2$.

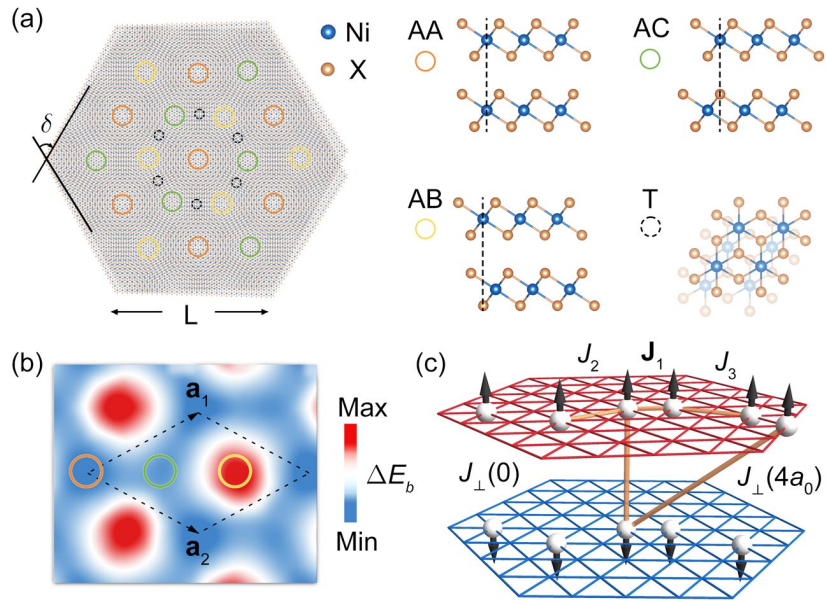

**Fig. 2**. (a) Crystal structure of twisted bilayer $NiX_2$ with twist angle $\theta = \delta - 60° = 1.096°$. Right panel in (a) shows the crystal structures of AA-, AB-, AC- and T-stacked patterns which are denoted by orange, yellow, green and black circles, respectively, in left panel. Blue and orange balls represent Ni and $X$ atoms, respectively. (b) Interlayer coupling energy $\Delta E_b$ of $NiCl_2$ as a function of an interlayer shift **r**. (c) Schematic diagram of intralayer and interlayer magnetic interactions in twisted bilayer $NiX_2$.

Building on the SL magnetic behavior, we extend our analysis to bilayer $NiX_2$ with interlayer exchange coupling quantified by interlayer coupling energy $\Delta E_b$, which is defined by the energy difference $\Delta E_b(\mathbf{r}) = E_{\text{FM}}(\mathbf{r}) - E_{\text{AFM}}(\mathbf{r})$ for lateral shifts $\mathbf{r} = \eta \mathbf{a}_1 + \nu \mathbf{a}_2$ ($\mathbf{a}_1$ and $\mathbf{a}_2$ are the lattice vectors). Computations of $\Delta E(\mathbf{r})$ on a 6 × 6 grids (validated against denser 9 × 9 and 11 × 11 grids; see **Fig. S4(c)** [45]) with $\eta, \nu = \left\{ 0, \frac{1}{6}, \cdots, \frac{5}{6} \right\}$ reveal spatially similar AFM-favored coupling ($\Delta E_b > 0$) for both bilayer NiCl$_2$ and NiBr$_2$ [see **Fig. 2(b)**], though coupling strength varies with ligand (**Table S1** [45]). From this resulting $\Delta E_b$, the interlayer exchange $J_\perp$ of bilayer Ni$X_2$ in **Eq. (7)** can be obtained, as detailed in Sec. III of Supplemental Material [45]). In contrast to the interlayer exchange that varies with stacking order, our DFT calculations indicate that the modifications to the intralayer interactions [isotropic exchange ($J_1, J_2, J_3$), magnetic anisotropy ($K$) and DMI ($\mathbf{J}_{1,ij}^A$)] imposed by interlayer coupling are negligible (**Table S2** [45]). We therefore construct the full bilayer spin Hamiltonian of **Eq. (1,4,7)** by combining the derived $J_\perp$ with magnetic parameters of SL Ni$X_2$. Based on this parameterized spin Hamiltonian, we perform atomistic spin simulations to investigate the spin textures of AA-stacked bilayer Ni$X_2$, which is energetically favored over the AB and AC stacking (**Table S3** [45]). As shown in **Figs. S3(d-f)** [45], AFM single-**q** spin-spiral features are observed in both cases. Crucially, bilayer NiCl$_2$ displays a longer spiral wavelength than NiBr$_2$, consistent with its weaker exchange frustration. This reduced frustration biases the spins toward $xy$-plane, enhancing the in-plane spin components in NiCl$_2$.

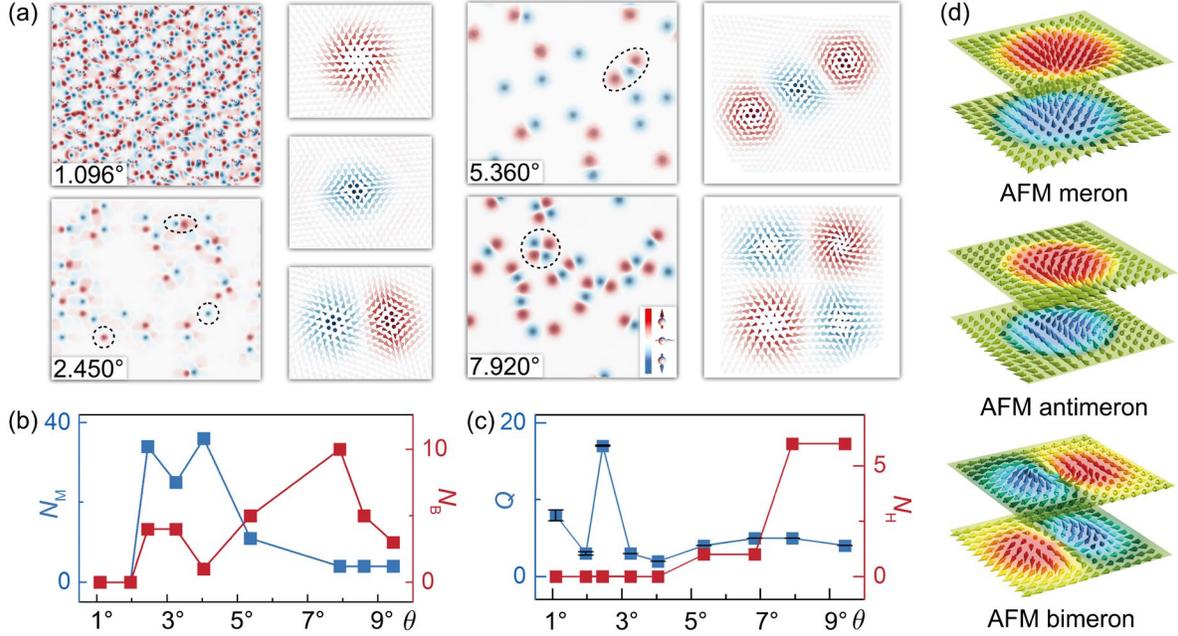

**Fig. 3**. (a) Spin textures of the top layer in twisted bilayer NiCl$_2$ at $\theta$ = 1.096°, 2.450°, 5.360° and 7.920°. Insets in (a) show the enlarged spin distributions of isolated merons, antimerons, bimerons, high-order merons and high-order bimerons. Out-of-plane spin components are denoted by colors and in-plane components are indicated by arrows. (b) Evolution of the total number of merons and antimerons, $N_M$, and bimerons, $N_B$, in twisted bilayer NiCl$_2$ as functions of $\theta$. (c) Evolution of the total number of high-order merons and high-order bimerons, $N_H$, and $Q_{up}$, in twisted bilayer NiCl$_2$ as functions of $\theta$. Error bars represent the deviation of $Q$ from the nearest integer. (d) Schematic illustrations of spin distributions for AFM topological magnetisms in twisted bilayer NiCl$_2$.

After establishing the AFM bilayer spin-spiral textures and uniform AFM interlayer coupling, we introduce an interlayer twist on the AA-stacked bilayer Ni$X_2$ (defined as $\theta = 0$) to form a long-period moiré pattern. The twist angle $\theta = \frac{180°}{\pi} \arccos\left[-\frac{1}{2} + \frac{3m(m-n)}{2(m^2 - mn + n^2)}\right]$ and moiré periodicity $L = a_0\sqrt{m^2 + n^2 + mn} = \frac{|m-n|a_0}{\sin(\theta_{m,n}/2)}$ are defined using a pair of coprime integers ($m$, $n$) to ensure the commensurability of moiré lattice. For simplicity, $n$ is fixed at 1 with $m$ variable. As **Fig. 2(a)** illustrates for $\theta = 1.096°$, the resulting long-period moiré lattice exhibits spatially varying stacking orders, mapped onto the high-symmetry configurations AA, AB, AC, and T with interlayer shifts $\mathbf{r} = 0, \frac{\mathbf{a}_1 + \mathbf{a}_2}{3}, \frac{2(\mathbf{a}_1 + \mathbf{a}_2)}{3}, \frac{\mathbf{a}_2}{2}$, respectively. This spatial variation generates tens of thousands of nonequivalent magnetic interactions within a moiré unit cell, complicating the determination of magnetic properties.

Following prior works [29-31], we extend the bilayer spin Hamiltonian to twisted bilayer Ni$X_2$ by adopting a widely employed approximation [40]: intralayer magnetic interactions are assumed identical to those in SL Ni$X_2$ and spatially uniform throughout the moiré unit cell, while the moiré interlayer coupling energy $\Delta E_m$ is derived from $\Delta E_b$ using a sliding-mapping approach [**Fig. S4(b)**]. This mapped moiré interlayer coupling energy $\Delta E_m$ allows us to determine the interlayer exchange interactions $J_\perp$ across the moiré lattice (see Sec. III of Supplemental Material [45]). With the first-principles parameterized moiré spin Hamiltonian established, we perform atomistic spin model simulations to explore the spin textures in twisted bilayer Ni$X_2$. It should be noted since we focus on the underlying physical mechanism, we restrict our analysis to the spin textures at $T = 0$ K. Given the reliability of our interlayer interaction approximation at small twist angle, we focus on the range of $\theta$ from 1.096° to 9.420°. **Figs. 3(a)** and **S5** [45] show the spin textures in the top layer of twisted bilayer NiCl$_2$ with different twist angle $\theta$. At $\theta = 1.096°$, the interplay between spin-spiral textures and moiré lattice gives rise to meron-antimeron loops, serving as an indication for topological magnetism [20,21]. Upon increasing $\theta$, these loops gradually shrink and eventually vanish by $\theta = 2.450°$, yielding isolated topological magnetisms on the top layer, including antimerons, merons and bimerons [**Fig. 3(a)**]. To quantify topological nature of these spin textures, topological number $Q = \frac{1}{4\pi}\iint \mathbf{S}(\mathbf{r}) \cdot [\frac{\partial \mathbf{S}(\mathbf{r})}{\partial x} \times \frac{\partial \mathbf{S}(\mathbf{r})}{\partial y}] d^2\mathbf{r}$ is calculated using the discrete form given in **Eq. S2**. **Fig. 3(c)** presents total topological number $Q_{up}$ of the top layer. For $\theta < 2.450°$, $Q_{up}$ is non-integer, indicating that spin textures are not fully composed of topological magnetic quasiparticles. In contrast, with $\theta \geqslant 2.450°$, the non-trivial nature of these spin textures is confirmed by the non-vanishing and quantized $Q_{up}$. In this regime, the emergent topological magnetism is revealed in the appearance of a multitude of isolated merons and antimerons, with only a few bimerons observed. Accordingly, the total number of isolated merons and antimerons, $N_M$, rises steeply, in contrast to the weak increase of the total number of isolated bimerons, $N_B$ [**Fig. 3(b)**]. As $\theta$ increases up to 5.360°, these isolated merons and antimerons begin to pair up. This pairing is evidenced by a sharp decline in $N_M$, accompanied by a surge in bimerons as well as the emergence of high-order merons [see **Figs. 3(b, c)**] [44]. With a further increase of $\theta$, isolated bimerons also start to coalesce, leading to a reduction in $N_B$ but the formation of high-order bimerons in the top layer. Turning from the top-layer viewpoint to the twisted bilayer as a whole, uniform interlayer AFM coupling stabilizes isolated AFM topological magnetisms [**Fig. 3(d)**] [8], with sublayers hosting spin textures with opposite topological numbers (**Fig. S6** [45]). This topological number compensation is expected to suppress the skyrmion Hall effect, a highly desirable feature for practical applications [43]. Despite the cancellation of the net topological number, unbalancing the layers provides a pathway for the detection and manipulation of these AFM topological quasiparticles [46]. Additionally, since lattice relaxation is known to occur in twisted bilayer $NiI_2$ [34, 35], we investigate its influence in twisted bilayer $NiCl_2$ and find that it has a minor impact on the emergence of topological magnetic quasiparticles (see Sec. VI in Supplementary Material). These results demonstrate the intriguing twist tunable moiré topological magnetism in twisted bilayer $NiCl_2$.

To get insight into the twist tunability of the moiré topological magnetism in twisted bilayer $NiCl_2$, we first explore the dependence of the spin textures on the stacking order. **Fig. S7(c)** [45] shows the spin textures of metastable AB- and AC-stacked bilayer $NiCl_2$. Spin-spiral textures are preserved in AA- and AC-stacked bilayer $NiCl_2$, while AB stacking favors an in-plane FM state. To pinpoint the origin of this dependence, evolutions of spin textures for bilayer $NiCl_2$ with the interlayer coupling energy $\Delta E_b$ are depicted in **Figs. S7(a, b)** [45]. Upon increasing the magnitude of $\Delta E_b$, spin spirals continuously transform into in-plane FM states. These results establish the interlayer coupling energy $\Delta E_b$ as the primary factor controlling spin textures in bilayer NiCl$_2$. In twisted bilayer NiCl$_2$, the stacking order varies across the moiré lattice, leading to a moiré modulation of interlayer coupling energy $\Delta E_b$. This modulation partitions the moiré unit cell into local regions, denoted by AA, AB, and AC (inset in **Fig. S8** [45]). At small twist angle, local regions are relatively large, and spin textures are governed locally by $\Delta E_b$. AA/AC regions favor meron–antimeron loops, whereas the AB region enforces in-plane textures. Further increasing the twist angle $\theta$ redistributes the sizes and arrangement of these $\Delta E$-defined local regions (see **Fig. S8** [45]), thereby endowing moiré topological magnetism with twist tunability. More details are shown in Sec. IV of Supplemental Material [45].

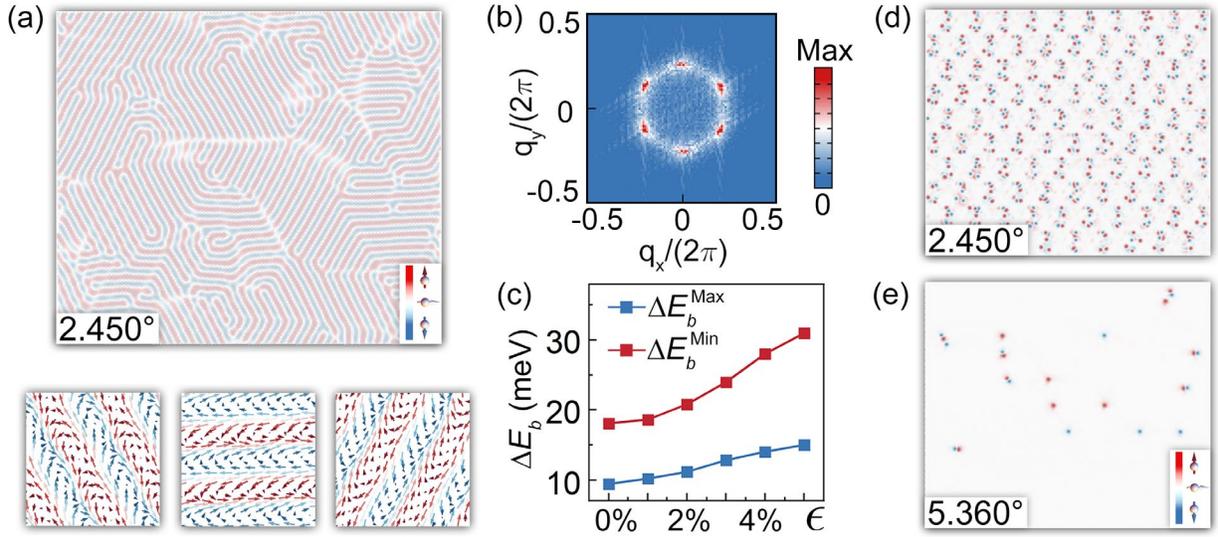

**Fig. 4**. (a) Spin textures of the top layer in twisted bilayer NiBr$_2$ at $\theta = 2.450°$. Insets in (a) show the enlarged spin distributions of three types of spin spirals in triple-**q** state. Out-of-plane spin components are denoted by colors and in-plane components are indicated by arrows. (b) Intensity plots of $S(\mathbf{q})$ in the extended Brillouin zone for spin textures in (a). (c) Evolution of interlayer coupling energy $\Delta E_b$ for bilayer NiBr$_2$ as a function of vertical compressive strain $\epsilon$. Red and blue dot lines represent $\Delta E_b^{Max}$ and $\Delta E_b^{Min}$, respectively. (d, e) Spin textures of the top layer in twisted bilayer NiBr$_2$ under $\epsilon = 5\%$ at $\theta = 2.450°$ and $5.360°$. Out-of-plane spin components are denoted

by color.

Unlike NiCl$_2$, the strong exchange frustration drives AFM spin-spiral textures in twisted bilayer NiBr$_2$ into a trivial AFM triple-**q** state, which is insensitive to twist angle $\theta$ (**Fig. S9** [45]). This triple-**q** state is composed of three degenerate spin spirals [insets in **Fig. 4(a)**], indicating trivial spin textures. Based on the above discussion, the interlayer coupling energy $\Delta E_b$ plays a crucial role in the emergence of topological magnetism. Since $\Delta E_b$ originates from interlayer interactions, it can be effectively tuned through applying vertical compressive strain $\epsilon$. Here, $\epsilon = \frac{r_s - r_0}{r_0} \times 100\%$ and $r_s$ represents the strained interlayer separation. As shown in **Fig. 4(c)**, both $\Delta E_b^{\text{Max}}$ and $\Delta E_b^{\text{Min}}$ increase with $\epsilon$, reflecting the enhanced strength of the interlayer coupling energy. We then investigate the influence of $\epsilon$ on spin textures in twist bilayer NiBr$_2$ at $\theta$ = 2.450°, as shown in **Figs. 4(d)** and **S10** [45]. Interestingly, at $\epsilon$ = 5%, twist bilayer NiBr$_2$ exhibits a coexistence of meron–antimeron loops and in-plane textures, indicating the emergence of twist-tunable topological magnetism. To confirm it, we explore the evolution of spin textures in strain twisted bilayer NiBr$_2$ with twist angle $\theta$. With increasing $\theta$ up to 5.360°, as shown in **Fig. 4(d)**, merons and antimerons pair up into bimerons, accompanied by the appearance of high-order merons, thereby establishing the twist tunable moiré topological magnetism in strained twisted bilayer NiBr$_2$.

## Conclusions

In summary, we establish a universal, field-free approach to generate and control moiré topological magnetism in 2D systems originally hosting trivial spin spirals. The frustration between twist-induced FM/AFM domains and uniform AFM interlayer exchange spontaneously stabilizes diverse topological magnetic phases, controlled purely by twist angle. First-principles and spin-model calculations reveal twist tunable isolated AFM bimerons, merons, antimerons, and high-order magnetic quasiparticles in NiCl$_2$, in contrast to trivial AFM spin-spiral textures in NiBr$_2$. These trivial spin spirals, however, can be transformed into intriguing topological magnetism through applying vertical compressive strain. Overall, this work establishes a new foundational platform to topological spintronics where topologically non-trivial spin textures can be generated from their trivial counterparts by 2D twist engineering.

## Conflict of Interest

The authors declare no competing financial interest.

## Acknowledgements

This work was supported by the National Natural Science Foundation of China (No. 12274261) and Taishan Young Scholar Program of Shandong Province. The research at University of Nebraska was supported by the National Science Foundation through the EPSCoR RII Track-1 program (NSF Grant No. OIA-2044049) and the UNL Grand Challenges catalyst award "Quantum Approaches Addressing Global Threats."